\documentclass[preprint,12pt]{elsart}
\usepackage{graphicx,amssymb,amsmath,times}
\journal{NRIAG Journal of Astronomy and Geophysics}
\begin{document}
\begin{frontmatter}
\title{Extragalactic Background Light models and GeV-TeV observation of blazars}
\author[label1,label2]{K. K. Singh\corauthref{cor}},
\corauth[cor]{Corresponding author.}
\ead{kksastro@barc.gov.in}
\author[label1]{P. J. Meintjes}
\address[label1]{Physics Department, University of the Free State, Bloemfontein- 9300, South Africa}
\address[label2]{Astrophysical  Sciences  Division, Bhabha Atomic Research Centre, Mumbai - 400 085, India}

\begin{abstract}
The extragalactic background light (EBL) in the ultraviolet to far-infrared wavelength range is dominated by 
the emissions from stars in galaxies and reflects the time-integrated history of the light production and 
reprocessing in the Universe. Direct measurements of the EBL are affected by the interplanetary dust 
and galactic emission. Hence, the absolute level of EBL is subject to considerable uncertainties. 
Observations of very high energy (VHE) blazars located at cosmological distances by the 
\textit{Fermi}-Large Area Telescope (LAT) and ground-based gamma-ray telescopes (e.g. H.E.S.S., MAGIC, VERITAS, TACTIC) 
provide a measurement of the EBL that is independent of the direct observations. The interaction of VHE or TeV photons 
originated from the distant blazars with the low energy EBL photons via $e^{-}e^{+}$ pair-production can be used as a 
powerful tool to probe the different EBL models in the wavelength range 0.1-1000$\mu$m. In this paper, we use two different 
methods to determine the opacity of the VHE $\gamma$-ray photons caused by the low energy EBL photons and 
study the consequences of six different EBL models. The first method- \emph{Model-Dependent Approach}, uses various EBL models 
for estimating the opacity as a function of the source redshift and energy. The second method- \emph{Model-Independent Approach}, 
relies on using the simultaneous observations of blazars in the MeV-GeV energy range from the \textit{Fermi}-LAT and in TeV band 
from the ground-based Cherenkov telescopes. We make underline assumption that the extrapolation of the LAT spectrum of blazars 
to TeV energies is either a good estimate or an upper limit for the intrinsic VHE spectrum of the source. We apply this method 
on the simultaneous observations of a few blazars PKS 2155-304, RGB J0710+591, 1ES 1218+304 and RBS 0413 at different redshifts 
to demonstrate a comparative study of six prominent EBL models. 
Opacities of the VHE $\gamma$-ray photons predicted by the \emph{Model}-Independent Approach are systematically larger 
than the ones estimated from the \emph{Model-Dependent Approach} using the six EBL models. Therefore, the $\gamma$-ray 
observations of blazars can be used to set a strict upper limit on the opacity of the Universe to the VHE $\gamma$-rays 
at a given redshift.
\end{abstract}
\end{frontmatter}
\section{Introduction}
The extragalactic background light (EBL) is the accumulated radiation from the structure formation and its cosmological
evolution. It consists of low energy photons emitted by stars and other cosmological objects at all epochs and is subsequently 
modified by redshifting and dilution due to the expansion of the Universe. The bulk of the EBL occurs at wavelengths from the 
optical-ultraviolet (UV) to the far-infrared (IR). The EBL photon field is dominated by the direct stellar emission in the 
optical to near-IR and by the stellar emission reprocessed by dust in the galaxies in the mid to far-IR [1]. Thus, the EBL 
is intimately connected with the star formation history of the Universe and reionization [2]. It provides very important 
information about the integrated star formation rate density and the cosmology. Therefore, absolute measurement of the EBL 
intensity is highly desirable. The EBL photons mostly lie in the wavelength range of 0.1-1000 $\mu$m and are assumed to 
be the second most energetic diffuse background in terms of the contained energy after the Cosmic Microwave Background 
Radiation (CMBR). Therefore, EBL has become essential for understanding the full energy balance of the Universe.
\par
Direct measurement of the EBL is a challenging task due to the strong foreground emission in our planetary system  and 
galaxy, some orders of magnitude larger than the actual EBL [3,4]. Direct measurements technically require absolute 
calibration of the instruments and understanding for the subtraction of all measurement uncertainties. 
Some direct measurements in optical [5] and in the near-IR [6,7] are available, but there is no general agreement 
about the reliability of these data from the observations [8]. The mid-IR band is known a little from the direct 
observations because of higher contamination from the zodical light of our Milkyway galaxy. 
Available observations provide lower limits on the density of EBL photons by using the integrated light from 
discrete extragalactic sources [9,10,11,12]. Semi-analytical modeling of the EBL density has also been 
performed by incorporating the simplified physical treatments of the key processes involved in the galaxy formation 
including gravitational collapse, merging of dark matter halos, gas cooling and dissipation, star formation, supernova 
feedback and metal production from the beginning of the Universe [13]. Modelling of the EBL leads to definite predictions, 
but uncertainties in the star formation rate, initial mass function, dust extinction and evolution with redshift have 
led to a significant discrepancy among various EBL models [14,15,16,17,18,19,20,21,22,23,24,25,26,27]. All the models 
have a limited predictive power for the EBL density, particularly as a function of time, because many details of the 
star and galaxy evolution remain unclear so far.
\par
Indirect measurements of the EBL photon density are possible from the observations of the very high energy 
(VHE; E $>$ 100 GeV) $\gamma$-ray emission from the distant sources. A beam of the VHE $\gamma$-ray photons  
traveling through the cosmological distances can be strongly attenuated by the  production of 
electron-positron (e$^-$-e$^+$) pairs in collisions with the low energy EBL photons [28]. Despite this effect, 
the current generation of the ground-based instruments (HESS, MAGIC, VERITAS, TACTIC) have significantly increased the number 
of observed extragalactic VHE $\gamma$-ray sources. The intrinsic VHE spectra of the extragalactic sources detected by 
the ground-based instruments depend on the spectrum of EBL, energy of the VHE photon as well as distance of the particular source. 
Blazars represent a very useful class of $\gamma$-ray beamers, being numerous over a wide range of redshifts, very luminous 
and long lasting sources. Despite a rigorous theoretical and observational studies in the literature, blazars are far from being 
standard candles, and therefore using them as tool for probing the EBL heavily depends on our understanding of their 
intrinsic emission and physical properties. Observation of the blazars by the large area telescope (LAT) onboard the 
\textit{Fermi} satellite in energy range from 30 MeV to more than 500 GeV provides a very strong observational evidence 
regarding the intrinsic MeV-GeV emission from them [29]. Combined MeV-GeV and TeV observations have provided 
a new tool to constrain the EBL intensity [30,31,32,33,34,35,36,37]. Recently, the \emph{Fermi}-LAT observations have been 
used to indirectly measure the EBL from the absorption features seen in the $\gamma$-ray spectra of blazars beyond 
the redshift $z >$ 1 [38,39]. Thus, TeV observation from the ground-based Cherenkov telescopes in association with 
the \emph{Fermi}-LAT observations can be used as powerful probe to study the EBL models. The details of current 
status of the direct and indirect measurements of the EBL can be found in [40,41,42,43].
\par
In this paper, we follow a methodology similar to one proposed by Georganopoulos et al. (2010) [44] to probe the 
six most recent and promising EBL models using simultaneous $\gamma$-ray observations of the four selected blazars 
at different redshifts from the \textit{Fermi}-LAT and ground-based instruments. The paper is organized as follows. 
Section 2 summarizes the six EBL models used in this study. In section 3, we report on the blazar observations and 
present understanding of the GeV-TeV emission. Section 4 describes the framework used in the present study. 
Application of this methodology and results are presented in Section 5 followed by discussion and conclusion in Section 6.
We have assumed a cosmology with parameters: $H_{0}$=70~km~s$^{-1}$~Mpc$^{-1}$, $\Omega_{m}$=0.30, $\Omega_{\Lambda}$=0.70 
under the framework of the flat $\Lambda$CDM geometry of the Universe.

\section{EBL Models}
The EBL is composed of stellar light emitted and partially reprocessed by the dust in the galaxies throughout the 
entire history of the cosmic evolution. Although, absolute level of the EBL density remains uncertain, the collective 
limits on the EBL from the direct and indirect measurements confirm the expected two peak structure in the spectral 
energy distribution (SED). The first hump lies in the UV-optical to near-IR wavelength range and peaks 
at $\lambda~ \sim$ 1$\mu$m. The second hump peaks at $\lambda~\sim$ 100$\mu$m in the far-IR regime. 
Due to the lack of direct EBL knowledge, many models have been presented in the last two decades. Based on the 
current understanding of the sources producing the EBL photons and their evolution in redshift, calculation of the EBL-SED 
is classified in four general categories [24]: \emph{forward evolution} (begins with the initial cosmological conditions 
and follows a forward evolution with time by means of the semi analytical models of the galaxy formation), 
\emph{backward evolution} (begins with existing  galaxy populations and extrapolates them backward in time), 
\emph{inferred evolution} (galaxy evolution is inferred from some observed quantity such as star formation rate density 
of the Universe over some range in wavelength), and \emph{observed evolution} (galaxy population is directly observed over 
a range of redshift that contributes significantly to the EBL). Some of the forward [13] and backward [15,19] evolution 
models have been disfavoured by the VHE observation of blazars. In the following, we briefly discuss six prominent EBL models 
which have been used in the present study.

\par\textbf{Franceschini et al. (2008) [20]:} It is a backward evolution model which provides estimates of the 
EBL photon density using available information on the cosmic sources producing diffuse photons in the Universe from far-UV to 
the sub-millimeter wavelengths over a wide range of the cosmic epochs with the best possible time and spectral 
resolution and their redshift evolution. This model exploits relevant data from the ground-based observatories in the 
optical, near-IR and sub-millimeter, as well as multi-wavelength information from the space-telescopes such as HST, 
ISO and Spitzer. Additional constraints are provided from direct measurements or upper limits on the EBL estimates 
by dedicated missions like COBE.

\par\textbf{Gilmore et al. (2009) [21]:} It is a forward evolution model based on semi-analytical models of the 
galaxy formation, which provides predictions of the dust extinguished UV radiation field due to the star-light 
and empirical estimates of the contributions due to the quasars. The model analyses predictions for the UV background 
that are intended to broadly span the possibilities in the star formation rate and quasar luminosity density. This 
model presents new calculations of the evolving UV component of the EBL out to the epoch of the cosmological 
reionization at high redshift.

\par\textbf{Finke et al. (2010) [22]:} It is an inferred evolution model for the UV through IR components of the EBL 
from the direct stellar radiation and reprocessed stellar radiation by the dust. This model takes into account the 
star formation rate, initial mass function, dust extinction and main-sequence stars as black bodies. The model is 
also extended to include the post-main sequence stars and reprocessing of starlight by the dust. The total energy 
absorbed by the dust is assumed to be re-emitted as three blackbodies in the IR, one at 40 K (warm, large dust grains), 
one at 70 K (hot, small dust grains) and one at 450 K (polycyclic aromatic hydrocarbons). This model does not require 
a complex stellar structure code or semi-analytical models of the galaxy formation.

\par\textbf{Kneiske et al. (2010) [23]:} It is also an inferred evolution model which produces a strict lower limit 
flux for the evolving EBL in the mid and near-IR range up to a redshift of $z=$ 5. A lower limit EBL model is 
derived by using the lower limit data from the integration of the galaxy number counts from the optical to far-IR region. 
The model takes into account the time-evolution of the galaxies, and includes the effect of the absorption and 
re-emission of the interstellar medium. The model is used to fit the observations of Spitzer, HST, ISO and GALEX to 
produce the complete EBL-SED.

\par\textbf{Dominguez et al. (2011) [24]:} It is an observed evolution model in which overall spectrum of the EBL 
between the wavelength range of 0.1-1000 $\mu$m is derived using a noble method based on the observations only. 
The method is based on the observed evolution of the rest frame K-band galaxy luminosity function up to a 
redshift of $z~\sim$ 4, combined with an estimation of the galaxy SED fractions. These quantities are achieved 
from fitting the Spitzer Wide Area Infrared Extragalactic Survey templates to a multi-wavelength survey sample of 
about 6000 galaxies in the redshift range of $z=$ 0.2-1 from the All-wavelength Extended Groth Strip International Survey. 
This model predicts EBL from UV to IR wavelength range and provides strong constraints on the EBL from UV to mid-IR, 
however, the far-IR component exhibits higher uncertainties.

\par\textbf{Gilmore et al. (2012) [25]:} It is a forward evolution model based on the latest semi-analytical models of the 
galaxy formation and evolution as well as an improved model for reprocessing of the star-light by the dust to mid and 
far-IR wavelengths. These semi-analytical models use a $\Lambda$CDM hierarchical structure formation scenario and 
successfully reproduce a large variety of the observational constraints on the galaxy number counts, luminosity and mass 
functions and color bi-modality. This model treats dust emission using empirical templates and predicts the EBL 
considerably lower than the optical and near-IR measurements.

\section{GeV-TeV $\gamma$-ray observations of Blazars}
Blazars are the most amazing class in active galactic nuclei (AGNs) family with a relativistic jet pointing towards
the line of sight of the observer at the Earth [45]. They are known for their broadband SED from radio to TeV-energy 
$\gamma$-rays and fast, large-scale variability in all bands. The present understanding of the blazars from 
observations points toward an SED with two spectral humps. The first hump peaks at the low energy from IR to X-ray 
and is assumed to be the synchrotron emission from a population of relativistic electrons in a partially 
ordered magnetic field. The second hump peaks at MeV-GeV energies and is thought to be the result of the 
inverse Compton scattering of the soft target synchrotron photons itself [46], photons from a dusty torus [47], 
photons from a broad-line region [48], or accretion disk photons [49] under the leptonic scenario. 
Alternative models associate the higher energy peak to the interaction of relativistic protons with an 
ambient photon field [50] or a hybrid population comprised of both leptons and hadrons [51]. 
For most of the blazars, it is believed that the TeV and GeV-emissions arise from the same physical mechanism and 
hence should be initimately related. Thus, the combined $\gamma$-ray observations of such blazars in the GeV and TeV 
regimes can be used to study the EBL in an indirect manner. The energy range of the \emph{Fermi}-LAT overlaps with 
the low energy threshold of the current generation ground-based $\gamma$-ray telescopes like MAGIC, H.E.S.S., VERITAS 
and TACTIC. Thus, for the first time, there is an excellent energy overlap between the space and ground-based 
telescopes, allowing simultaneous observations of the continuous spectra between 100 MeV and 20 TeV produced from blazars. 
Blazars, that are most likely to be detected by both, the \textit{Fermi}-LAT and the ground-based telescopes simultaneously, 
are therefore crucial for this study. 

\section{Framework}
Observation of the VHE $\gamma$-rays from blazars at the cosmological distances can be used as an alternative and 
completely independent way with respect to direct measurements to probe the EBL. The approach is based on the 
study of the absorption features imprinted on the GeV-TeV spectra due to the interaction of $\gamma$-rays with the 
EBL photons. Measurement of the effect of suppression of the TeV $\gamma$-ray emission from the blazars at non-negligible 
level can provide an estimate of the EBL density in the local Universe. Two completely different phenomenological 
approaches for studying the EBL density predicted by various models (Section 2) are described below in detail.

\subsection{Model-Dependent Approach}
In this method, we study the effect of EBL on the propagation of VHE $\gamma$-ray photons traveling through intergalactic 
space from sources at known redshifts. A very important consequence of the EBL is the attenuation of the VHE  $\gamma$-rays 
emitted by the sources at cosmological redshifts through electron-positron pair creation [28]. The physical process involved 
is expressed as
\begin{equation}
	\gamma_{VHE} + \gamma_{EBL} \rightarrow e^{-} + e^{+}
\end{equation}
From the theory of radiation transfer, above process gives rise to an exponential decay of the intrinsic 
VHE $\gamma$-ray flux emitted from the distant blazars. The observed VHE $\gamma$-ray flux on earth is related 
to the intrinsic flux of the source as
\begin{equation}
	F_{obs}= F_{int} \times e^{-\tau (E,z_s)}
\end{equation}  
The optical depth $\tau (E,z_s)$ encountered by the VHE $\gamma$-rays of energy $E$ emitted from the source at 
redshift $z_s$ and traveling towards the Earth due to the EBL absorption is given by
\begin{equation}
	\tau(E,z_s)=\int\limits_{0}^{z_s}\left(\frac{dl}{dz}\right)dz
	              \int\limits_{0}^{2}\frac{\mu}{2} d\mu
                      \int\limits_{\varepsilon_{th}}^{\infty}n_{EBL}(\varepsilon,z)
		      \sigma_{\gamma\gamma}(E,\varepsilon,\mu)d\varepsilon
\end{equation} 
where $\mu= 1-cos \theta$, $\theta$ being angle between the momenta of two photons in the lab frame, 
$\epsilon$ is the energy of EBL photon undergoing pair production with VHE $\gamma$-ray photon, 
and $n_{EBL} (\varepsilon,z)$ is the EBL photon number density. 
Threshold energy of the EBL photons for pair production is given by
\begin{equation}
	\varepsilon_{th}(E,\mu,z)=\frac{2m_{e}^{2}c^{4}}{E \mu (1+z)^2}
\end{equation}
where $m_{e}$ is the rest mass of electron. $\sigma_{\gamma\gamma}(E,\varepsilon,\mu)$ is the total cross-section 
for pair creation and is defined as [52]
\begin{equation}
 \sigma_{\gamma\gamma}(E,\varepsilon,\mu)=\frac{\pi r_{0}^{2}}{2}(1-\beta^{2})\left[(3-\beta^{4})\rm ln\frac{1+\beta}{1-\beta}-2\beta(2-\beta^{2})\right]
\end{equation}
where, the Lorentz factor $\beta$ represents the velocity of $e^{-}$ or $e^{+}$ in the center of mass system and it depends on 
$E$, $\varepsilon$ and $\theta$. $r_{0}$ is the classical electron radius. For an isotropic distribution of the 
low energy EBL photons, the pair production cross-section has a distinct peak close to the threshold corresponding to 
$\beta~\approx$ 0.70. This implies that the cross-section is maximized for the EBL photon interacting with a 
VHE $\gamma$-ray photon provided following condition is satisfied
\begin{equation}
	\lambda_{EBL}(\mu m)=1.187\times (1 + z)^2 \times E (TeV)
\end{equation}
Hence, VHE $\gamma$-rays at a rest frame energy above 1 TeV are most likely absorbed by the mid and far-IR 
range of the  EBL photons, while those in the 100 GeV to 1 TeV regime are sensitive to the EBL photons in the near-IR 
and optical bands. Below 100 GeV, it is mainly UV part of EBL-SED that causes the attenuation. Below 20 GeV, there is 
little absorption due to the increasing scarcity of the hard UV background photons. Thus, the attenuation of the 
VHE $\gamma$-ray photons by the EBL can in principle be used to estimate the EBL density at wavelengths corresponding 
to the observations of $\gamma$-rays from blazars at cosmological redshifts. The line element for a $\gamma$-ray photon 
moving from source to observer in the $\Lambda$CDM cosmology is expressed as
\begin{equation}
	\frac{dl}{dz}=\frac{c}{H_0}\frac{1}{(1+z)\sqrt{\Omega_{\Lambda} + \Omega_{m}(1+z)^3}}
\end{equation}

Besides redshifting all energies in proportion of $(1+z)$ for cosmological applications, the cosmic expansion dilutes 
the EBL density by a factor $(1+z)^{3}$. In addition, EBL spectral energy distribution changes because of the 
intrinsic evolution of the galactic population over cosmic times. The EBL photons are progressively produced 
by the galaxies, but their density builds up slowly through the star formation history of the Universe. Therefore, the photon 
comoving number density decreases with redshift and is lower in the case of expanding Universe than that in the static Universe.
Also, optical and near-IR photons are produced at lower $z$ than the far-IR photons and therefore their comoving number 
density decreases faster with redshift. Various methods followed to model the evolution of EBL photon density 
suggest that $n_{EBL} (\varepsilon,z)$ acquires an extra factor and dilutes as $(1+z)^{3-k}$ in the expanding 
Universe [53,54,55,56]. The value of evolution factor $k$ lies between 1.1 to 1.8 for $z <$ 1. For our present calculations, 
we assume $k=$ 1.2 as it shows good agreement between different approaches [31,55]. 
The EBL photon spectral number density, which depends on the adopted model for the EBL, is a key ingredient in the 
evaluation of the optical depth. This is obtained from the SED predicted by different EBL models using the 
following conversion factor,
\begin{equation}
 n(\varepsilon)[cm^{-3} eV^{-1}]= 1.70395\times10^{-4}\times \lambda^{2}[\mu m]\nu I_{\nu}[nW m^{-2} sr^{-1}]
\end{equation}
Using the above methodology, optical depth of the VHE $\gamma$-ray photons for a given EBL model can be determined 
as a function of $z_s$ and $E$. Hence, we refer this method as the \textit{Model-Dependent Approach}.

\subsection{Model-Independent Approach}
In this method, we use near simultaneous observations of the blazars with the \emph{Fermi}-LAT and ground-based 
$\gamma$-ray telescopes. With the launch of \emph{Fermi}-LAT, the MeV-GeV observations of the blazars are now 
possible in a regime where the EBL attenuation is negligible. Overlapping of the operational energy range of 
the \emph{Fermi}-LAT and ground-based $\gamma$-ray telescopes makes the blazar observations an important 
tool to probe the opacity of the Universe to VHE $\gamma$-rays as they propagate from their sources to the Earth. 
The Universe appears to be largely transparent to $\gamma$-rays at all the \emph{Fermi}-LAT energies and 
out to redshift $z \sim$ 2, whereas opaque to the TeV photons at $z \le$ 0.2. We assume that, in the \emph{Fermi}-LAT 
operational energy range (0.1-500 GeV) blazar spectra are good representation of the intrinsic spectra and 
extrapolation of the \emph{Fermi}-LAT spectra in the VHE range (0.1-20 TeV) gives the intrinsic TeV spectra of the blazars. 
Therefore, if the intrinsic spectra of blazars are described by any spectral form in GeV-TeV regime without need for a break, 
their observed spectra would be imprinted with a break solely attributed to the EBL absorption. From the theory of 
radiative transfer, optical depth for a given blazar is given by
\begin{equation}
	\tau(E,z_{s})=\rm ln\left(\frac{F_{1}}{F_{2}}\right) \pm \left[\left(\frac{\Delta F_{1}}{F_{1}}\right)^{2} + 
		      \left(\frac{\Delta F_{2}}{F_{2}}\right)^{2}\right]^{\frac{1}{2}}
\end{equation}
where ($F_{1} \pm \Delta F_{1}$) is the \emph{Fermi}-LAT flux extrapolated to the VHE range and ($F_{2} \pm \Delta F_{2}$) 
is the corresponding TeV flux measured by the ground-based instruments. Thus, using the above expression, optical depth of 
the VHE photons can be estimated using the \emph{Fermi}-LAT (MeV-GeV) and ground-based (TeV) observations of the blazars 
at a given $z_s$ without using any EBL model. Hence, we refer this method as the \textit{Model-Independent Approach}. 
We apply this methodology to a few selected blazars as discussed below.

\section{Results}
We use the methodology described in Section 4 to a few selected blazars PKS 2155-304, RGB J0710+591, 1ES 1218+304 
and RBS 0413 at different redshifts to study the six EBL models. The underlying assumption in the present study 
is that the \emph{Fermi}-LAT operational energy range is practically unaffected due to the EBL absorption and 
the spectra of blazars in the LAT energy range represent a correct measure of the intrinsic spectra in the VHE regime. 
Extrapolation of the MeV-GeV spectrum to the TeV energy range gives a strict upper limit on the intrinsic TeV flux of the 
source. The TeV fluxes measured by the ground-based telescopes carry the imprint of the EBL absorption in the blazar spectra. 
We briefly discuss below the \emph{Fermi}-LAT and TeV-observations of the above four blazars and their use in the present study 
in the order of increasing redshifts. 

\subsection{PKS 2155-304}
\begin{figure}
\begin{center}
\includegraphics*[height=0.60\textheight,angle=-90]{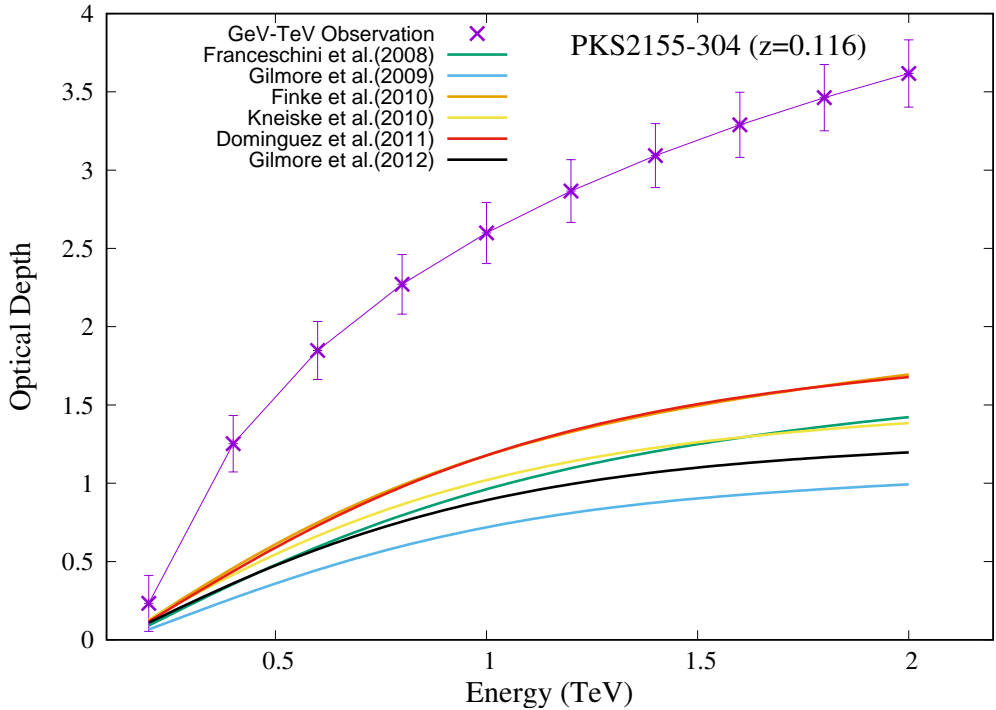}
\caption{Comparison of the optical depth values estimated using two different methods: 
	\emph{Model-Dependent Approach} (six EBL) and \emph{Model-Independent Approach} (GeV-TeV Observations) 
	for $z_s =$ 0.116.} 
\label{fig:pks2155}
\end{center}
\end{figure}

PKS 2155-304 is a high frequency peaked BL Lac (HBL) type of blazar at redshift $z_s =$ 0.116. It was simultaneously 
observed by the \textit{Fermi}-LAT and H.E.S.S. telescopes in 2008 [57]. The time averaged \emph{Fermi}-LAT and 
VHE spectra of PKS 2155-304 are well described by a power-law with photon spectral indices of 
$\Gamma_{1}$= 1.81$\pm$0.11 and $\Gamma_{2}$= 3.34$\pm$0.05 respectively [57]. Using these measurements, 
we have estimated the optical depth of VHE photons emitted from the source in the energy range 0.2-2 TeV by using 
the \emph{Model-Independent Approach}. We use six different EBL models to calculate the optical depth values in the 
energy range 0.2-2 TeV at redshift $z=$ 0.116 using the \emph{Model-Dependent Approach}. The optical depth values 
obtained from two different approaches for $z_s=$ 0.116 are shown in  Figure \ref{fig:pks2155} and are also reported 
in Table \ref{tab:pks2155} (Appendix). The blazar PKS 2155-304 was observed in nearly quiescent state without 
any signature of variability in the MeV-GeV and TeV light curves. Therefore, the intrinsic VHE spectrum of 
the source can be modified only due to the EBL absorption. Hence, the \emph{Model-Independent Approach} predicts 
the maximum opacity of the Universe to VHE $\gamma$-rays at $z_s =$0.116. We observe that the optical depth values 
derived from the GeV-TeV observations in the energy range 0.2-2 TeV are larger than that from the \emph{Model-Dependent Approach}. 
\subsection{RGB J0710+591}
\begin{figure}
\begin{center}
\includegraphics*[height=0.60\textheight,angle=-90]{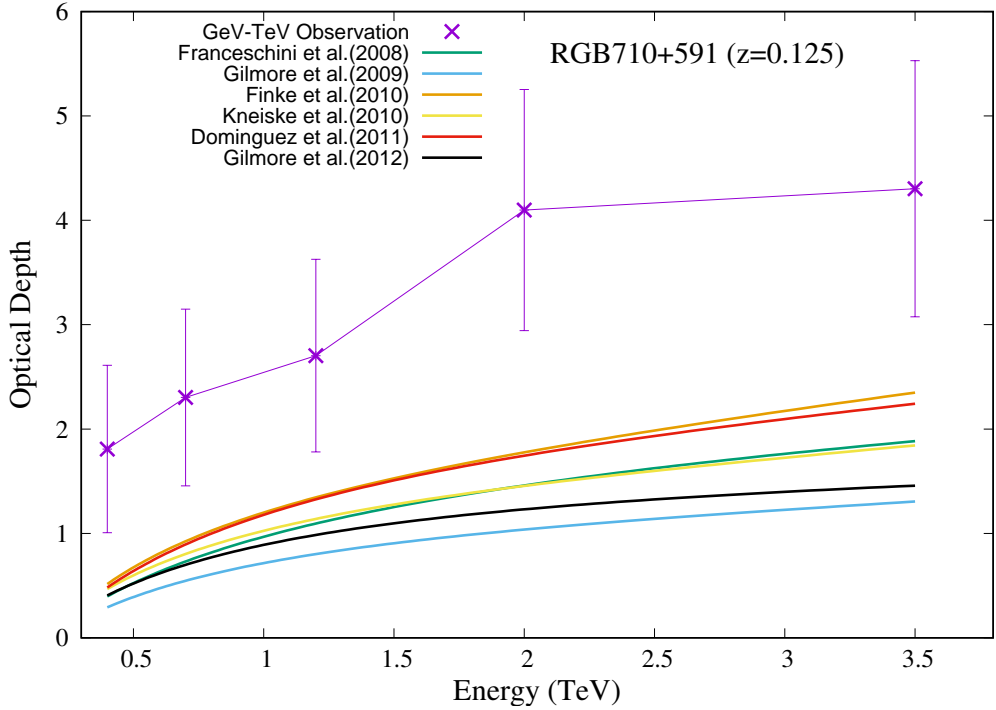}
\caption{Same as Figure 1 for $z_s =$ 0.125.} 
\label{fig:rgb710}
\end{center}
\end{figure}

RGB J0710+591 is a well known extreme blazar featuring in many catalogs at redshift $z_s =$ 0.125. 
The first VHE $\gamma$-ray emission from this source was discovered by the VERITAS array of telecopes 
during 2008-09 observations [58]. The VHE observations were complemented by contemporeneous observations 
with the \emph{Fermi}-LAT. The time averaged spectra in the MeV-GeV and TeV energy bands are fitted by a simple 
power-law with photon spectral indices of $\Gamma_{1}$= 1.46$\pm$0.17 and $\Gamma_{2}$= 2.69$\pm$0.26 respectively [58].
From these observations, we have calculated the opacity of TeV photons in the energy range 0.4-3.5 TeV by using the 
\emph{Model-Independent Approach}. We also use the six EBL models to determine the opacity in the above energy range 
at source redshift $z_s =$ 0.125 following the \emph{Model-Dependent Approach}. The optical depth values from two different 
approaches are depicted in Figure \ref{fig:rgb710} and are also summarized in Table \ref{tab:rgb710} (Appendix) for comparison. 
It is obvious from the figure that GeV-TeV $\gamma$-ray observations experience more opacity in the intergalactic space 
than that provided by the six different EBL models at $z_s =$ 0.125. The large error bars in the values of optical depth 
obatined from the \emph{Model-Independent Approach} are attributed to the uncertainties in the observed VHE spectra. 
Due its relatively hard $\gamma$-ray spectra and no evidence of variability [59], RGB J0710+591 is one the most 
suitable blazars to probe the EBL models.

\subsection{1ES 1218+304}
\begin{figure}
\begin{center}
\includegraphics*[height=0.60\textheight,angle=-90]{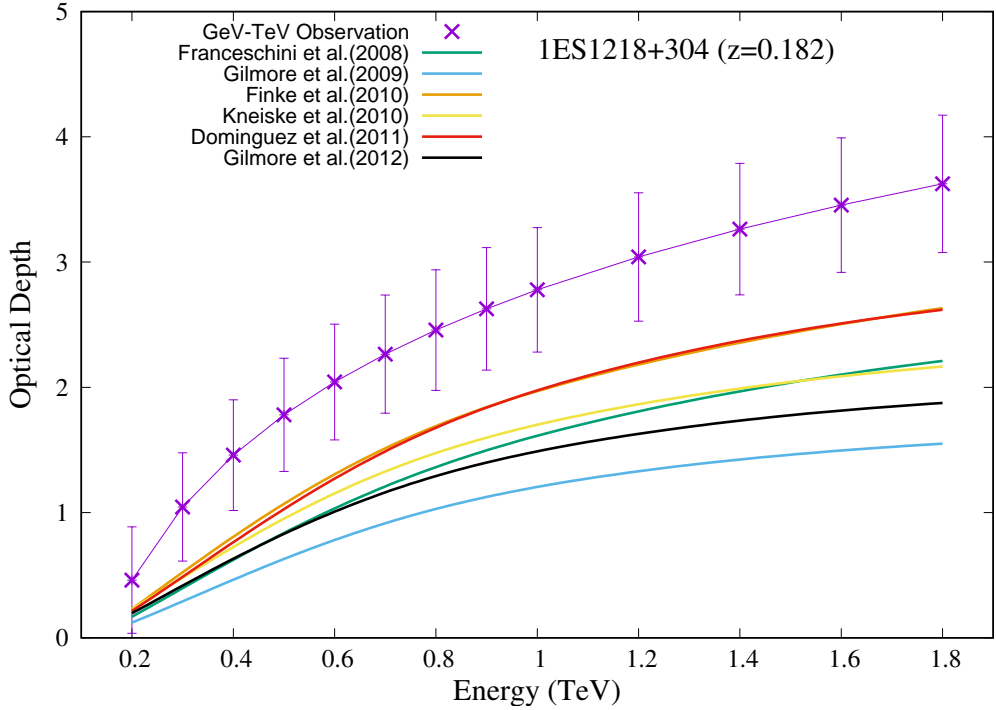}
\caption{Same as Figure 1 for $z_s =$ 0.182.} 
\label{fig:1es1218}
\end{center}
\end{figure}

The HBL object 1ES 1218+304 at redshift $z_s =$ 0.182 belongs to a group of blazars that exhibit unusually hard VHE spectra 
considering their redhsifts [60,61]. This blazar was observed by the VERITAS telescope from December 2008 to May 2009 and 
the time averaged VHE spectrum was described by a power-law with a photon spectral index of $\Gamma_{2}$=3.07$\pm$0.09 [61].
The quasi-simultaneous \emph{Fermi}-LAT spectrum is also described by a power-law with a photon spectral index of 
$\Gamma_{1}$= 1.63$\pm$0.12 making it one of the hardest $\gamma$-ray source. From the GeV-TeV quasi-simultaneous observations 
of the blazar 1ES 1218+304, we have estimated the optical depth of the VHE $\gamma$-ray photons coming from the source 
in the energy range 0.2-1.8 TeV using the \emph{Model-Independent Approach}. Going to the higher redshift $z_s =$ 0.182, 
we calculate the optical depth in the energy range 0.2--1.8 TeV corresponding to the six EBL models by applying 
the \emph{Model Dependent Approach}. The two estimates of optical depths at $z_s =$ 0.182 are shown in 
Figure \ref{fig:1es1218} and are also reported in Table \ref{tab:1es1218} (Appendix). We clearly observe that the 
\emph{Model-Independent Approach} again predicts the highest opacity as compared to the \emph{Model-Dependent Approach}. 

\subsection{RBS 0413}
\begin{figure}
\begin{center}
\includegraphics*[height=0.60\textheight,angle=-90]{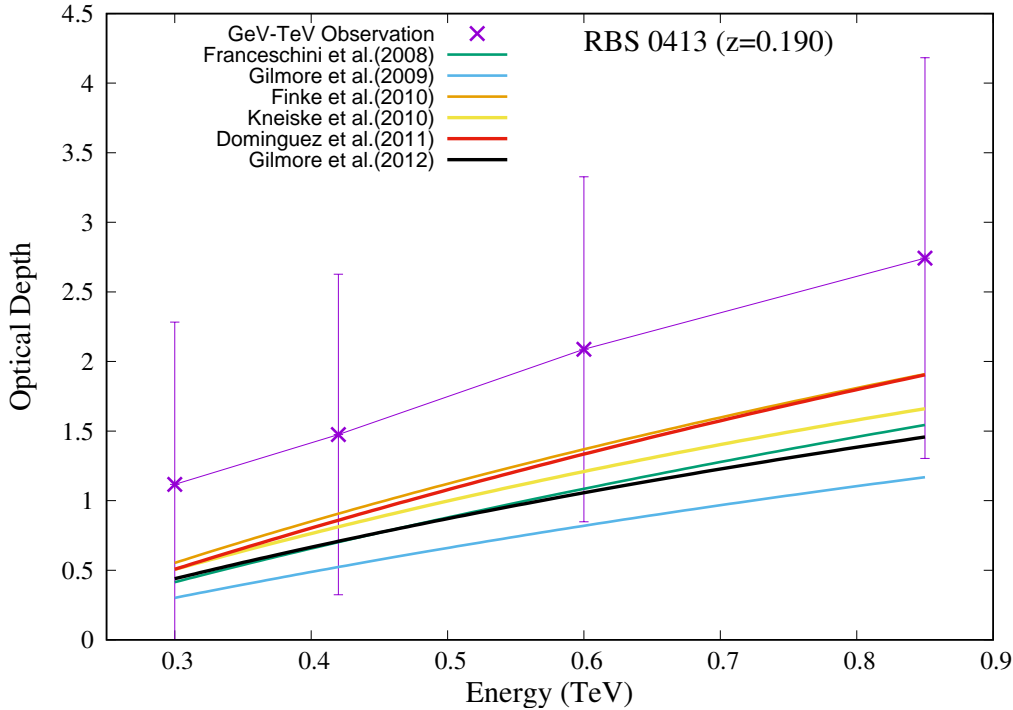}
\caption{Same as Figure 1 for $z_s =$ 0.190.} 
\label{fig:rbs0413}
\end{center}
\end{figure}

RBS 0413 was discovered in X-ray band during the Einstein Medium Sensitivity Survey and was later identified as an HBL 
located at redshift $z_s =$ 0.190. It is a weak source in the VHE regime. The VHE emission from this source was detected 
by the VERITAS telescope and was also complemented by the contemporaneous observation with the \emph{Fermi}-LAT [62]. 
The observed VHE spectrum can be described by a power-law with a photon spectral index of $\Gamma_{2}$= 3.18$\pm$0.68 and 
the MeV-GeV spectrum from the \emph{Fermi}-LAT observations has a photon spectral index of $\Gamma_{1}$= 1.57$\pm$0.12. 
Using these two contemporaneous observations, we have estimated the optical depth values in the energy range 0.3-0.85 TeV 
from the \emph{Model-Independent Approach}. The optical depths in the same energy band have also been calculated using the 
six EBL models under the framework of the \emph{Model-Dependent Approach} at $z_s =$ 0.190. These values of the optical depths 
obtained from two different approaches are compared in Figure \ref{fig:rbs0413} and have also been given in 
Table \ref{tab:rbs0413} (Appendix). 
The large error bars in the optical depth values obtained from the \emph{Fermi}-LAT and VHE observations are due to the 
higher uncertainties in the flux extrapolation from the LAT energy range to the TeV energies. Despite large error bars, 
the GeV-TeV observations predict more opacity than any EBL model used in the present study at redshift $z_s=$ 0.190. 

\section{Discussion and Conclusion}
We have used two distinct and completely independent approaches to study the opacity of the Universe to the VHE $\gamma$-ray
photons emitted at different redshifts. The \emph{Model-Independent Approach} uses simultaneous observations of 
four selected blazars by the \emph{Fermi}-LAT and ground-based $\gamma$-ray telescopes. The \emph{Fermi}-LAT observations 
are used as a proxy for the intrinsic source emission in the GeV energy regime. The intrinsic TeV spectra of blazars are 
obtained by a simple extrapolation of the \textit{Fermi}-LAT spectra to the VHE energies. The VHE spectra of blazars 
measured by the ground-based \emph{TeV} instruments are expected to suffer EBL absorption of the VHE $\gamma$-ray photons. 
The EBL photons in different wavebands affect each part of the blazar spectrum in a different way. Over some energy band 
like MeV-GeV and GeV-TeV, the spectra of most of the blazars can be approximated by a simple power-law shapes. That is, 
if the intrinsic spectrum is a power-law, the observed spectrum with the EBL absorption can also be described by a power 
law with steeper spectral index. The amount of steepening in the VHE spectra gives an indirect estimate of the EBL absorption 
of the \emph{TeV} photons. From the present study, we conclude the following:

\begin{itemize}

\item Using the \emph{Fermi}-LAT and TeV-observations of the selected blazar spectra, we have estimated the opacity 
of the Universe to the VHE $\gamma$-rays at different redshifts $z_s =$ 0.116, 0.125, 0.182 \& 0.190. We compare 
these estimates with the values calculated for six different EBL models using the \emph{Model-Dependent Approach}.

\item A comparison of the two results indicates that the \emph{Model-Independent Approach} which completely 
relies on the blazar observations predicts the highest opacity of the Universe to the VHE $\gamma$-rays at 
all the redshifts, and it is beyond the values predicted by any EBL model used in this study. 
This implies that the EBL level predicted by any of the six models is not sufficient to produce the opacity of the Universe 
to the VHE $\gamma$-rays as expected by the GeV-TeV observations of the blazars.

\item Since, we assume that the extrapolation of the \emph{Fermi}-LAT spectra to the TeV energies gives maximum 
level of the intrinsic TeV flux, the opacities obtained from the \emph{Fermi}-LAT and \emph{TeV}-observations of the 
blazars using the \textit{Model-Independent Approach} can be used as upper limits for the opacity of the Universe. 
Therefore, any EBL model which gives opacity of the Universe through the \textit{Model-Dependent Approach} higher than 
that predicted by the blazar observations can be excluded or disfavoured. 
However, it is important to mention here that the \textit{Model-Independent Approach} discussed in this study is based 
on the assumption that the \emph{Fermi}-LAT and VHE spectra are simultaneously measured in the quiescent state, when no 
temporal and spectral intrinsic variability are present in the source. The $\gamma$-ray spectra measured during the 
orphan flares or transient events from the blazars are not suitable for the \textit{Model-Independent Approach}. 
Also, the presence of a curvature in the intrinsic combined GeV-TeV spectra of the blazars can predict higher values of the 
opacity using the \textit{Model-Independent Approach}. In that case, a simple extrapolation of the \emph{Fermi}-LAT spectra 
to the TeV energy range may not be a good description of the intrinsic $\gamma$-ray spectra of the blazars. 

\item We attribute any steepening of the blazar spectral slope other than EBL absorption intrinsic to the source. 
For better prediction of the intrinsic TeV spectra, the local opacity caused by absorption taking place within the source 
can be incorporated while modelling the spectral energy distribution of blazars, but this is beyond the scope of this work.

\item Among the six EBL models used in the present study, Gilmore et al. (2009) [21] predicts the lowest opacity, 
whereas models proposed by Finke et al. (2010) [22] and Dominguez et al. (2011) [24] give similar and highest 
opacity of the Universe to the VHE $\gamma$-rays at different redshifts. This indicates that Finke et al. (2010) [22] 
and Dominguez et al. (2011) [24] EBL models predict similar EBL-SED and show better consistency with the GeV-TeV 
observations of the blazars. Also, Franceschini et al. (2008) [20] and Kneiske et al. (2010) [23] EBL models 
represent similar EBL intensity but show less consistency with the opacity expected from the blazar observations. 
The EBL models proposed by Gilmore et al. (2009 \& 2012) [21,25] significantly differ from all other models used 
in the present study and both models give less opacity to the VHE $\gamma$-rays than remaining EBL models.
This suggests that the \emph{Model-Independent Approach} based on $\gamma$-ray observations either over estimates 
the opacity of the Universe to the VHE $\gamma$-rays or the EBL models predict lower values of the optical depth at 
different redshifts. The too low values of the optical depth obtained from the \emph{Model-Dependent Approach} 
can be attributed mainly to the lack of exact measurement of the EBL-SED and less understanding of the 
proper cosmological evolution of the EBL density in the models. The optical depth values corresponding to 
the Finke et al. (2010) [22] and Dominguez et al. (2011) [24] can be scaled up by a factor $\sim$ 2 to get the 
better agreement with the \emph{Model-Independent Approach} within statistical uncertainties at all four redshifts 
considered in this study. Recently, Desai et al. (2019) have used a sample of 38 blazars to measure the EBL intensity 
using the \emph{Fermi}-LAT and ground-based observations [35]. The optical depth at the GeV energies estimated from the 
from the \emph{Fermi}-LAT observations [39] and TeV optical depths from the multiple spectra of 38 blazars in 
the energy range 0.1-21 TeV are combined with normalized opacities from the known EBL models to constrain the EBL 
intensity.
\end{itemize} 

As the \emph{Fermi}-LAT continuously monitors the $\gamma$-ray sky, the improved simultaneous measurements of the GeV-TeV spectra 
of more blazars over a range of redshift with the current generation ground-based instruments such as MAGIC, H.E.S.S., VERITAS, 
and TACTIC can provide a deep understanding of the existing EBL models using the approach discussed in this paper, which is 
similar to the one used by Georganopoulos et al. (2010) [44]. In future, the improved sensitivity of the Cherenkov Telescope 
Array (CTA) and its wide energy coverage will help in better understanding of the EBL [63]. With the quick accumulation of the 
blazars observed by the \emph{Fermi}-LAT and ground-based $\gamma$-ray telescopes, which will be enehanced by the future CTA, 
the \emph{Model-Independent Approach} can be used to measure the EBL by studying the absorption imprint in the spectra 
of number of blazars with greater accuracy. Alternative scenarios to explain the VHE spectra of distant blazars 
and EBL absorption do exist in the literature. The instrinsic VHE spectra of blazars from the broadband SED modelling are 
also used as a probe for EBL models [64]. Axion like particle (ALPs), which had been proposed to explain the strong-CP violation 
problem in the particle physics, could lead to a conversion of the VHE $\gamma$-ray photons into ALPs in the presence of intergalactic magnetic 
field [65,66]. This photon-ALP conversion drastically reduces the EBL absorption effects and enlarges the VHE $\gamma$-ray horizon. 
Production of secondary $\gamma$-rays along the line of sight by the interactions of cosmic-ray protons with the background photons 
have also been used to explain the VHE spectra of blazars at cosmological distances [67]. 
\section*{Acknowledgement}
We thank the anonymous reviewers for their important and helpful suggestions to improve the contents of this study.
\section*{Appendix (Tables 1-4)}
Optical depth estimates from the \emph{Model-Indepenedent} (GeV-TeV observations) and \emph{Model-Dependent}  (EBL models) 
methods at four redshifts. The six EBL models are designated as, Fran-2008: Franceschini et al. (2008), Gil-2009: Gilmore et al. (2009), 
Kneis-2010: Kneiske et al. (2010), Fink-2010: Finke et al. (2010), Domin-2011: Dominguez et al. (2011), and Gil-2012: Gilmore et al. (2012).
\begin{table}
\caption{$z=$ 0.116 (PKS 2155-304)}
\begin{center}
\begin{tabular}{lcccccccc}
\\	
\hline
E	&Model-Independent	&\multicolumn{6}{c}{Model-Dependent}\\
(TeV)	&		&Fran-2008&Gil-2009&Kneis-2010&Fink-2010&Domin-2011&Gil-2012\\
\hline
	0.20	&0.2326$\pm$0.1785	&0.0912	&0.0658	&0.1255	&0.1287	&0.1179	&0.1089\\		
	0.40	&1.2526$\pm$0.1801	&0.3628	&0.2667	&0.4708	&0.4292	&0.4407	&0.3714\\
	0.60	&1.8476$\pm$0.1851	&0.6291	&0.4769	&0.8023	&0.7137	&0.7738	&0.6204\\
	0.80	&2.2705$\pm$0.1904	&0.8397	&0.6383	&1.0533	&0.9182	&1.0369	&0.8035\\
	1.00	&2.5985$\pm$0.1954	&1.0040	&0.7569	&1.2285	&1.0673	&1.2358	&0.9371\\
	1.20	&2.8665$\pm$0.2000	&1.1273	&0.8407	&1.3629	&1.1763	&1.3795	&1.0263\\
	1.40	&3.0931$\pm$0.2043	&1.2245	&0.8951	&1.4673	&1.2478	&1.4837	&1.0889\\
	1.60	&3.2894$\pm$0.2082	&1.3012	&0.9318	&1.5475	&1.3021	&1.5634	&1.1375\\
	1.80	&3.4626$\pm$0.2119	&1.3680	&0.9652	&1.6267	&1.3496	&1.6247	&1.1682\\
	2.00	&3.6174$\pm$0.2152	&1.4225	&0.9929	&1.6948	&1.3841	&1.6783	&1.1972\\	
\hline
\end{tabular}
\end{center}
\label{tab:pks2155}
\end{table}

\begin{table}
\caption{$z=$ 0.125 (RGB J0710+591)}
\begin{center}
\begin{tabular}{lcccccccc}
\\
\hline
E	&Model-Independent	&\multicolumn{6}{c}{Model-Dependent}\\
(TeV)	&		&Fran-2008&Gil-2009&Kneis-2010&Fink-2010&Domin-2011&Gil-2012\\
\hline
	0.40	&1.8084$\pm$0.8019	&0.3970	&0.2917	&0.5152	&0.4682	&0.4821	&0.4059\\
	0.70	&2.3026$\pm$0.8465	&0.8067	&0.6142	&1.0413	&0.8958	&0.9952	&0.7818\\
	1.20	&2.7035$\pm$0.9230	&1.2220	&0.9105	&1.4772	&1.2739	&1.4946	&1.1108\\
	2.00	&4.0982$\pm$1.1548	&1.5388	&1.0739	&1.8341	&1.4965	&1.8158	&1.2937\\
	3.5	&4.3023$\pm$1.2269	&1.8840	&1.3055	&2.3495	&1.8432	&2.2427	&1.4579\\
\hline
\end{tabular}
\end{center}
\label{tab:rgb710}
\end{table}

\begin{table}
\caption{$z=$ 0.182 (1ES 1218+304)}
\begin{center}
\begin{tabular}{lcccccccc}
\\	
\hline
E	&Model-Independent	&\multicolumn{6}{c}{Model-Dependent}\\
(TeV)	&		&Fran-2008&Gil-2009&Kneis-2010&Fink-2010&Domin-2011&Gil-2012\\
\hline
	0.20	&0.4606$\pm$0.4252	&0.1683	&0.1212	&0.2307	&0.2306	&0.2149	&0.1962\\
	0.30	&1.0445$\pm$0.4319	&0.3909	&0.2845	&0.5218	&0.4777	&0.4784	&0.4148\\
	0.40	&1.4588$\pm$0.4418	&0.6328	&0.4670	&0.8206	&0.7366	&0.7691	&0.6434\\
	0.50	&1.7801$\pm$0.4521	&0.8599	&0.6476	&1.1082	&0.9833	&1.0543	&0.8576\\
	0.60	&2.0426$\pm$0.4621	&1.0645	&0.8088	&1.3476	&1.1983	&1.3115	&1.0430\\
	0.70	&2.2646$\pm$0.4717	&1.2432	&0.9455	&1.5611	&1.3702	&1.5343	&1.1982\\
	0.80	&2.4569$\pm$0.4808	&1.3977	&1.0619	&1.7396	&1.5186	&1.7264	&1.3306\\
	0.90	&2.6265$\pm$0.4893	&1.5352	&1.1592	&1.8896	&1.6400	&1.8915	&1.4405\\
	1.00	&2.7782$\pm$0.4973	&1.6534	&1.2418	&2.0135	&1.7458	&2.0311	&1.5306\\
	1.20	&3.0408$\pm$0.5121	&1.8442	&1.3628	&2.2222	&1.9041	&2.2464	&1.6614\\
	1.40	&3.2628$\pm$0.5253	&1.9902	&1.4417	&2.3768	&2.0140	&2.4030	&1.7564\\
	1.60	&3.4550$\pm$0.5373	&2.1131	&1.5016	&2.5123	&2.0951	&2.5212	&1.8223\\
	1.80	&3.6246$\pm$0.5483	&2.2101	&1.5502	&2.6327	&2.1662	&2.6187	&1.8749\\
\hline
\end{tabular}
\end{center}
\label{tab:1es1218}
\end{table}

\begin{table}
\caption{$z=$ 0.190 (RBS 0413)}
\begin{center}
\begin{tabular}{lcccccccc}
\\
\hline
E	&Model-Independent	&\multicolumn{6}{c}{Model-Dependent}\\
(TeV)	&		&Fran-2008&Gil-2009&Kneis-2010&Fink-2010&Domin-2011&Gil-2012\\
\hline
	0.30	&1.1164$\pm$1.1669	&0.4142	&0.3015	&0.5525	&0.5058	&0.5067	&0.4388\\
	0.42	&1.4753$\pm$1.1516	&0.7175	&0.5329	&0.9288	&0.8298	&0.8743	&0.7264\\
	0.60	&2.0877$\pm$1.2399	&1.1210	&0.8518	&1.4190	&1.2606	&1.3812	&1.0972\\
	0.85	&2.7430$\pm$1.4400	&1.5437	&1.1684	&1.9089	&1.6607	&1.9038	&1.4572\\		
\hline
\end{tabular}
\end{center}
\label{tab:rbs0413}
\end{table}


\end{document}